\input harvmac

%
%
%

\def\bar{\overline}
\def\hat{\widehat}
\def\*{\star}
\def\[{\left[}
\def\]{\right]}
\def\({\left(}		
\def\){\right)}

%
%
\def\zb{{\bar{z} }}
\def\frac#1#2{{#1 \over #2}}
\def\inv#1{{1 \over #1}}

\def\d{\partial}

\def\2pi{\hbox{$2\pi i$}}

\def\dsl{\raise.15ex\hbox{/}\kern-.57em\partial}
\def\Dsl{\,\raise.15ex\hbox{/}\mkern-.13.5mu D}
%
%

	\def\Sig{\Sigma}

%
%

%

\def\2pi{\hbox{$2\pi i$}}

\def\dsl{\raise.15ex\hbox{/}\kern-.57em\partial}
\def\Dsl{\,\raise.15ex\hbox{/}\mkern-.13.5mu D}
%
%
%
\font\numbers=cmss12
\font\upright=cmu10 scaled\magstep1
\def\stroke{\vrule height8pt width0.4pt depth-0.1pt}
\def\topfleck{\vrule height8pt width0.5pt depth-5.9pt}
\def\botfleck{\vrule height2pt width0.5pt depth0.1pt}
\def\Zmath{\vcenter{\hbox{\numbers\rlap{\rlap{Z}\kern
0.8pt\topfleck}\kern
2.2pt
                   \rlap Z\kern 6pt\botfleck\kern 1pt}}}
\def\Qmath{\vcenter{\hbox{\upright\rlap{\rlap{Q}\kern
                   3.8pt\stroke}\phantom{Q}}}}
\def\Nmath{\vcenter{\hbox{\upright\rlap{I}\kern 1.7pt N}}}
\def\Cmath{\vcenter{\hbox{\upright\rlap{\rlap{C}\kern
                   3.8pt\stroke}\phantom{C}}}}
\def\Rmath{\vcenter{\hbox{\upright\rlap{I}\kern 1.7pt R}}}
\def\Z{\ifmmode\Zmath\else$\Zmath$\fi}
\def\Q{\ifmmode\Qmath\else$\Qmath$\fi}
\def\N{\ifmmode\Nmath\else$\Nmath$\fi}
\def\C{\ifmmode\Cmath\else$\Cmath$\fi}
\def\R{\ifmmode\Rmath\else$\Rmath$\fi}


\Title{NSF-ITP-97-021,  hep-th/9703055}
{\vbox{\centerline{Errata for: Differential Equations}
\centerline{ for Sine-Gordon Correlation Functions   }
\centerline{ at the Free Fermion Point
} 
}}

\bigskip
\bigskip

\centerline{Denis Bernard\foot{Member of CNRS} }
\medskip
\centerline{Service de Physique Th\'eorique, CEN-Saclay\foot{Laboratoire
de la Direction des sciences de la mati\`ere du Commissariat \`a
l'\'energie atomique.} }
\centerline{F-91191 Gif sur Yvette, France}
\bigskip

\centerline{Andr\'e Leclair\foot{On leave from Cornell University}} 
\medskip
\centerline{Institute for Theoretical Physics}
\centerline{University of California}
\centerline{Santa Barbara, CA 93106-4030}

\vskip .3in

We present some important corrections to our work which
appeared in Nucl. Phys. B426 (1994) 534.  Our previous
results for the correlation functions 
$\langle e^{i\alpha \Phi(x)} e^{i\alpha' \Phi (0) } \rangle$
were only valid for $\alpha = \alpha'$,  due to the fact 
that we didn't find the most general solution to the
differential equations we derived.  Here we present the
solution corresponding to $\alpha \neq \alpha'$.      

\Date{2/97}
%
%
%
%
%
%
%
%
%
%
%
%
%
%
%
%

\appendix{E}{Errata}

In section 3 we did not find the most general solution to 
the differential equations (3.37) when we imposed 
$\d_z a = \d_z b = \d_\zb (b-a) = 0$.  We now understand
that for $\alpha \neq \alpha'$, the latter condition is not
valid.  In this errata we present the modifications 
for $\alpha \neq \alpha'$.
A corrected version of the paper which incorporates the
modifications below is available\ref\rus{D. Bernard and
A. Leclair, hep-th/9402144.}\foot{Recently, similar results 
were obtained using different
methods in \ref\rwidom{H. Widom, {\it An Integral Operator Solution
to the Matrix Toda Equations}, solv-int 9702007.}. } 

\bigskip

(1)  ~~Equation 1.3 should be replaced with:
\eqn\result{\eqalign{
\( \d_r^2 + \inv{r} \d_r \)   \Sigma (r ) & =  \frac{m^2}{2}
\( 1 - \cosh 2\varphi \) \cr
\( \d_r^2 + \inv{r} \d_r \)  \varphi &= \frac{m^2}{2} \sinh 2\varphi
+ \frac{4 (\alpha - \alpha')^2 }{r^2} \tanh \varphi (1-\tanh^2 \varphi )
, \cr}}
where $r^2 = 4 z \zb$, and $m$ is the mass....

\bigskip

(2)  ~~In equation (3.31), $\d_\zb B_+ = \frac{m}{2} \hat{C}_+ C_- $
should be replaced with $\d_\zb \hat{B}_+ = \frac{m}{2} \hat{C}_+ C_- $.

\bigskip

(3)  ~~The end of section 3, beginning with the sentence after (3.38),
should be replaced with the following:

Inserting  this parameterization  into the differential equations
gives the following.  The first two equations in (3.37)  give
\eqn\emodi{
\d_z a = -\tanh^2 \varphi ~ \d_z b . }
Using this equation and its $\d_\zb$ derivative the second 2 equations
can be simplified to
\eqn\emodii{\eqalign{
( \d_z \d_\zb a )  ~ \coth \varphi  - ( \d_z \d_\zb b ) ~\tanh \varphi
-2 \d_z \varphi ~ \d_\zb (b-a ) =0 \cr
\d_z \d_\zb \varphi = \frac{m^2}{2} \sinh 2\varphi
- \tanh \varphi ~ \d_z b ~ \d_\zb (b-a) . \cr }}
 
The function $b$ can be deduced using Lorentz invariance.
Let $z = r e^{i \theta} /2 $, $\zb = r e^{-i \theta} /2 $, and
consider shifts of $\theta$ by $\gamma$.
The functions $e, \hat{e}$ satisfy
\eqn\emodiii{\eqalign{
e( e^{i\gamma} z , e^{-i \gamma} \zb , u )
&= e^{-i\gamma ( 1 + \alpha' - \alpha )/2 } e( z, \zb , e^{i\gamma} u ) \cr
\hat{e} ( e^{i\gamma} z , e^{-i \gamma} \zb , u )
&= e^{-i\gamma ( 1 + \alpha - \alpha' )/2 } e( z, \zb , e^{i\gamma} u ) 
. \cr }}
From the definition (3.21)  of $C_+, \hat{C_+}$, and making the change 
of variables $u \to e^{-i\gamma} u$, one finds 
\eqn\emodmod{
C_+ = e^{2i(\alpha - \alpha') \theta}  f(r) ,
~~~~~
\hat{C}_+ = e^{-2i(\alpha - \alpha') \theta}  \hat{f}(r) ,
}
for some scalar functions $f , \hat{f}$.   Then, using
\eqn\emodxx{
e(z, \zb , u) = u ~ \hat{e} (\zb, z, 1/u ) , }
one can show $f = \hat{f}$ by making the change of variables $u\to 1/u$.
Thus,
\eqn\emodiv{
e^{2b} = \frac{C_+}{\hat{C}_+} =
e^{4 i (\alpha - \alpha' ) \theta}
, } and
\eqn\emodv{ 
b = (\alpha - \alpha') \log \( \frac{z}{\zb} \) . }
Inserting this $b$ into \emodi\ and taking the complex conjugate, one
deduces
\eqn\emodvi{
\d_\zb a = \tanh^2 \varphi ~ \d_\zb b . }
The function $\varphi$ is only a function of $r$.  Thus
\emodii\ can be written as
\eqn\resulti{
\( \d_r^2 + \inv{r} \d_r \)  \varphi = \frac{m^2}{2} \sinh 2\varphi
+ \frac{4 (\alpha - \alpha')^2 }{r^2} \tanh \varphi (1-\tanh^2 \varphi )
. }

Finally, using the equation (3.36), and also (3.38), one obtains
\eqn\EDx{
\( \d_r^2 + \inv{r} \d_r \)  \Sig(r) = - m^2\ \sinh^2\varphi
=  \frac{m^2}{2} \({1 - \cosh 2\varphi   }\) . }
This is the result announced in the introduction.
Notice that $\d_z \d_\zb \Sig$ is only parameterized by a single
function $\varphi (r)$, and the differential equation for
$\varphi$ involves only $\varphi$ itself.

\bigskip

We thank S. Lukyanov for first pointing out a possible error in
the original paper.  The result \result\ was used in the work
\ref\rluk{S. Lukyanov and A. Zamolodchikov, {\it Exact expectation
values of local fields in quantum sine-Gordon model}, 
hep-th/9611238.  }. 

\bigskip

This work is supported by the National Science foundation, in part
through the National Young Investigator Program, and under 
Grant No. PHY94-07194.

\listrefs
\end